# Insulator-metal transition and the magnetic phase diagram of La$_{1-x}$Te$_x$MnO$_3$ $(0.1 \leq x \leq 0.6)$

J. Yang[1], W. H. Song[1], Y. Q. Ma[1], R. L. Zhang[1], B. C. Zhao[1], Z. G. Sheng[1], G. H. Zheng[1], J. M. Dai[1], and Y. P. Sun*[1,2]

[1]*Key Laboratory of Internal Friction and Defects in Solids, Institute of Solid State Physics, Chinese Academy of Sciences, Hefei, 230031, P. R. China*

[2]*Laboratory of Solid State Microstructures, Nanjing University, Nanjing 210008, P. R. China*

**Abstract**

The structural, electrical transport and magnetic properties of perovskite oxides La$_{1-x}$Te$_x$MnO$_3$ $(0.1 \leq x \leq 0.6)$ have been investigated and thus the magnetic phase diagram of La$_{1-x}$Te$_x$MnO$_3$ $(0.1 \leq x \leq 0.6)$ compounds as a function of temperature and the doping level x has been obtained. All samples have rhombohedral structure and undergo paramagnetic-ferromagnetic (PM-FM) transition accompanied with metal-insulator transition (MIT). Whereas a charge ordering (CO) transition begins to appear at $T_{CO} \sim 250K$ for the sample with x=0.60. Moreover, the variation of the Curie temperature $T_C$ and the MIT temperature $T_P$ is quite complex and the results are discussed in terms of three factors including the average A-site cation radius <r$_A$>, the size mismatch $\sigma^2$ and the Te content. In addition, there has an evident magnetoresistance (MR) at low temperatures for all samples.

Corresponding author.   Phone: 86-551-5592757; Fax: 86-551-5591434; email: *ypsun@issp.ac.cn*

# 1. Introduction

Recently, doped manganese perovskites have attracted much renewed attention due to the discovery of colossal magnetoresistance (CMR) [1]. In the systems of $Ln_{1-x}A_xMnO_3$ (Ln=La-Tb, and A=Ca, Sr, Ba, Pb, etc.), the existence of the dopant leads to mixed valent of the Mn ions ($Mn^{3+}/Mn^{4+}$). Many theories have been proposed to explain the mechanism about CMR such as double exchange [2], polaronic effects [3], phase separation combined with percolation [4,5] and Griffiths singularity [6]. Their studies indeed suggested that the mixed valent of $Mn^{3+}/Mn^{4+}$ is a key component for understanding the CMR effect and the transition from the ferromagnetic metal to paramagnetic semiconductor. Note that there exists an inherent symmetry between $Mn^{2+}$ and $Mn^{4+}$ as both are non Jahn-Teller ions whereas $Mn^{3+}$ is a Jahn-Teller ion. So electron doping is expected to drive the manganese into a mixture of $Mn^{2+}$ and $Mn^{3+}$ when using the tetravalent element to dope the parent compound $LnMnO_3$. Thus, the basic physics in terms of Hund's rule coupling and Jahn-Teller effect could operate in the electron doped phase as well. Recently, many researches have placed emphasis on electron-doped compound such as $La_{1-x}Ce_xMnO_3$ [7-10], $La_{1-x}Zr_xMnO_3$ [11] and $La_{2.3-x}Y_xCa_{0.7}Mn_2O_7$ [12]. Their studies suggested that the CMR behavior probably occurred in the mixed-valent state of $Mn^{2+}/Mn^{3+}$.

In the following we report Te doping manganites $La_{1-x}Te_xMnO_3 (0.1 \leq x \leq 0.6)$. This material has been reported by Tan et al. in the doping range of 0.04-0.2 [13-15]. Their studies demonstrated that the Curie temperature $T_C$ and the MIT temperature $T_P$ monotonously increase with increasing Te content. However, our experimental data indeed shows that the variation of $T_C$ and $T_P$ is not monotonously in the broad range of 0.1-0.6. Moreover, the magnetic phase diagram of $La_{1-x}Te_xMnO_3$ $(0.1 \leq x \leq 0.6)$ as a function of temperature and the doping level x was obtained from the magnetic and resistivity measurements. In order to understand the change of $T_C$ and $T_P$ clearly, we combine the structure and electronic transport properties with the A-site cation mean radius $<r_A>$, the size mismatch $\sigma^2$ and the characteristic temperature $T_0$.

## 2. Experiment

A series of ceramic samples of $La_{1-x}Te_xMnO_3 (0.1 \leq x \leq 0.6)$ were synthesized by a conventional solid-state reaction method in air. The powders mixed in stoichiometric compositions of high-purity $La_2O_3$, $TeO_2$ and $MnO_2$ were ground, then fired in air at 700 °C for 24h. The powders obtained were ground, palletized, and sintered at 1030 °C for 24h with three intermediate grinding, and finally, the furnace was cooled down to room temperature. The structure and lattice constant were determined by powder x-ray diffraction (XRD) using $CuK_\alpha$ radiation at room temperature. The resistance as a function of temperature was measured by the standard four-probe method from 25 to 300K. The magnetic measurements were performed on a Quantum Design superconducting quantum interference device (SQUID) MPMS system ($2 \leq T \leq 400$ K, $0 \leq H \leq 5$ T).

## 3. Results and discussion

The magnetic phase diagram of $La_{1-x}Te_xMnO_3 (0.1 \leq x \leq 0.6)$ as a function of temperature and the doping level x is shown in Fig.1. The curved line is drawn as a guide to the eye approximately on the boundary between the spin-disordered paramagnetic state and the spin-ordered state. The structural, magnetic and transport properties of the manganites $La_{1-x}Te_xMnO_3$ in the various doping regimes are discussed below.

### 3.1. Structural characterizations

X-ray powder diffraction (XRD) at room temperature shows that the samples are single phase below x=0.2 and the second phase begin to exist beyond x=0.2, which is consistent with many electron-doped manganites [8,11]. But the amount of second phase is below 5%, we can consider that the samples are almost single phase. XRD patterns of all samples have a rhombohedral lattice with the space group $R\bar{3}C$. The structural parameters are refined by the standard Rietveld technique [16], and the experimental spectra and calculated XRD pattern for x=0.1 are shown in Fig.2. The structural parameters obtained are listed in Table I. As we can see, the Mn-O-Mn bond angle increases initially, and then decreases with increasing Te content. Especially there exist a minimum in Mn-O-Mn bond angle for x=0.25. In succession, the bond angle increases again and stabilizes a definite value for higher Te content. Whereas the Mn-O bonds length displays the inverse correlation to the variation in the Mn-O-Mn bond angle.

3.2. Magnetic properties

The Magnetization of $La_{1-x}Te_xMnO_3$ $(0.1 \leq x \leq 0.6)$ under 0.1T magnetic field cooling (FC) is measured. Fig.3 (a) shows the dependence of M on temperature T. The $T_C$ (defined as the one corresponding to the peak of $dM/dT$ in the M vs. T curve) values are obtained as 235, 236, 188, 184, 185, 233, 177, 181, 182 and 186K for x=0.10, 0.15, 0.20, 0.25, 0.30, 0.35, 0.40, 0.45, 0.50 and 0.60, respectively. The Curie temperature $T_C$ increases slightly when the x value changes from x=0.1 to x=0.15, decreases abruptly from x=0.15 to x=0.30 but stabilizes for $0.20 \leq x \leq 0.30$, then $T_C$ increases rapidly when x value changes from x=0.30 to x=0.35 and decreases sharply but varies steadily for higher values. The saturation of $T_C$ may be attributed to a competition between the double exchange and the core spin interaction, which leaded to the canting of the core spins as the doped level increases [11]. As for the variation of $T_C$, we will discuss in the following. Moreover, there exist a little anomaly seen at a temperature of about 40K in the M-T curve. In the previous study [14], similar phenomenon appears in the M-T curve, but the authors neglect this and ascribed the character of bifurcation at low temperature in the FC and ZFC curves to the appearance of the spin-glass. In order to clarify the magnetic structure of the samples, the frequency dependence of the ac magnetic susceptibility ($\chi$) for the samples with x=0.20 and 0.40 was performed under the frequency of 188, 1172, and 1876 Hz, respectively. As shown in Fig.3 (b), upon cooling from room temperature the in –phase χ′(T) curves display abrupt upturns at about 200K and 180K for the samples with x=0.20 and 0.40, respectively, exhibiting the PM-FM transition. Upon further decreasing the temperature, the χ′(T) curves display a cusp at about 40K for the two samples. Moreover, the peak temperature of the cusp does not shift towards higher temperature with increasing frequency. These features are characteristic of phase transition. To our best knowledge, the Curie temperature of $Mn_3O_4$ is about 41K [17,18]. So we suggest that the second phase $Mn_3O_4$ provides a small contribution to the magnetization of the samples at about 40K and the amount of impurity increases with increasing the doping level corresponding with the distinctness of the anomaly for higher doping level.

### 3.3. Transport properties

The temperature dependence of the resistivity of $La_{1-x}Te_xMnO_3 (0.1 \leq x \leq 0.5)$ is shown in Fig.4 (a). The experimental data were obtained in 0 and 0.5T magnetic fields for all the samples. As we can see, there is an insulator-metal transition caused by paramagnetic-ferromagnetic phase transition under zero and 0.5T magnetic fields for all the samples. Moreover, when external magnetic field is applied, the resistivity is suppressed significantly and the temperature of resistivity peak shifts slightly to a higher value. This suggests that the external magnetic field facilitates the hopping of $e_g$ between the neighboring Mn ions, which agrees with the DE model. At the same time, the magnitude of the resistivity vary several orders with the amount of dopant Te from x=0.10 to x=0.60. Only one resistivity peak is observed for the samples with $x \geq 0.20$, whereas for the samples with x=0.10 and x=0.15, an additional shoulder can be observed. The resistivity peak at the high temperature corresponding to the metal-insulator transition is defined as $T_{P1}$ and the temperature of the low-temperature transition peak is denoted by $T_{P2}$ (221K for x=0.10 and 222K for x=0.15). The values of $T_{P1}$ for the studied samples obtained from the $\rho$ (T) curve are 243, 246, 224, 136, 204, 216, 211, 185 and 221K for x=0.10, 0.15, 0.20, 0.25, 0.30,

0.35, 0.40, 0.45 and 0.50, respectively. As we can see, the variation of $T_{P1}$ and $T_C$ are very alike. In addition, similar phenomenon with double peak in the $\rho$ - T curve has also been observed in alkaline-earth-metal-doped and alkali-metal-doped samples of LaMnO$_3$ [19-21]. However, its real origin is not very clear until now. Zhang et al. [21] considered that the peaks at $T_{P2}$ are believed to reflect the spin-dependent interfacial tunneling due to the difference in magnetic order between surface and core. Ilryong Kim et al. [22] ascribe the low temperature resistivity peak to the electroneutral type phase separation (PS) induced by non-uniformly distributed oxygen. Also there were many other explanations for the origin of the low temperature resistivity peak. Here we attributed the resistivity peak at $T_{P2}$ to the grain boundaries (GBs).

Fig.4 (b) shows the resistivity as a function of temperature under 0 and 0.5T magnetic fields for the sample with x=0.60. As we can see, upon cooling from room temperature, the sample shows a charge ordering (CO) transition at $T_{CO} \sim 250K$ and the CO state was suppressed greatly under the magnetic field. Upon further decreasing temperature, the metal-insulator transition is observed at $T_P$=206K. It is worth noting that the resistivity behavior at low temperature. The resistivity does not decrease with decreasing temperature, but begin to increase at about $T^*$=68K. In addition, the magnetization as a function of the applied magnetic field at 5K is shown in the inset. For the sample with x=0.60, the magnetization M (H) at low magnetic fields resembles a long-range ferromagnetic ordering, whereas the magnetization M increases slightly without saturation at higher fields, which reveals that the antiferromagnetic (AFM) phase give rise to the linear high-field region. As it is well known, the ferromagnetic DE interaction favors $e_g$ electrons transfer between neighboring Mn ions, leading to a transition to FM metallic behavior at $T < T_P$. On the other hand, the AFM superexchange interaction prevents $e_g$ electrons to transfer between neighboring Mn ions, leading to an increase in resistance at low temperatures.

So it can be concluded that the sample with x=0.6 does not show homogeneous FM phase at low temperature, but shows the characteristic of coexistence of the FM phase and AFM phase, and the competition between them lead to yield a minimum at about 68K in the ρ vs. T curve.

In order to understand the electronic transport mechanism of the studied samples clearly, the resistivity data above $T_P$ for all samples are fitted by the thermally activated (TA) law [ $\rho \propto \exp(E_0/k_B T)$ ], the adiabatic small polaron hopping (SPH) model [ $\rho \propto T\exp(E_P/k_B T)$ ] and the variable range hopping model (VRH) [$\rho \sim \exp(T_0/T)^{1/4}$], respectively [23]. The results show that $\rho(T)$ curves can be well described by VRH model as shown in Fig.5. Moreover, $T_0$ can be achieved from the variable-range hopping fitting data. It is well known that $T_0$ is related to the spatial extension $(l)$ of the localized states and to the density of states $[g(E_F)]$: $l \approx [g(E_F)T_0]^{-1/3}$. It is clear that rising of the $T_0$ value when increasing the bending of the Mn-O-Mn bond should reflect the enhancement of the carrier effective mass or the narrowing of the bandwidth [24]. The $T_0$ values are also listed in Table I. Our data suggested that the increase of the $T_0$ values associated to the decrease of the localization length reduces the carrier mobility and thus the resistivity increases accompanied with $T_P$ shifting towards lower temperatures, which have been confirmed by many researches.

In order to investigate the variation of $T_P$ and $T_C$ with the amount of dopant Te further, we calculated the average A-site cation radius <$r_A$>, the variance [25] (second moment) of the A-cation radius distribution $\sigma^2 = \sum y_i r_i^2 - <r_A>^2$ (where $y_i$ is the fractional occupancy of A-site ion and $r_i$ is the corresponding ionic radii). Standard ionic radii [26] with values 1.216Å for $La^{3+}$, and 0.97Å for $Te^{4+}$, respectively, were used to calculate $<r_A>$ and $\sigma^2$. It is well known that for hole-doped manganites, the following three factors have been shown to strongly affect the DE (and hence $T_C$ and $T_P$), i.e., the hole carriers density controlled by the $Mn^{3+}/Mn^{4+}$ ratio, the average A-site cation size $<r_A>$ and the A-site size mismatch effect $\sigma^2$ [25, 27-31]. From the point view of being favorable to stabilize the low-temperature FM metallic phase one would expect an optimum $Mn^{3+}/Mn^{4+}$ ratio to be 2:1. On the other hand, the optimum $Mn^{3+}/Mn^{4+}$ ratio is favorable to form an ideal cubic perovskite. Any deviation from the ideal cubic perovskite would lead to a reduction in the Mn-O-Mn bond angle from 180°, which directly weakens the DE. Beside the $Mn^{3+}/Mn^{4+}$ ratio, both $<r_A>$ and $\sigma^2$ have also been shown to influence the DE. The principal effect of decreasing $<r_A>$ is to decrease the Mn-O-Mn bond angle, thereby reducing the matrix element b that described electron hopping between Mn sites [27], while the mismatch effect would promote the localization of $e_g$ electrons thereby weakening the DE. Unlike hole-doped manganites, the mixed-valence of manganese ions will be $Mn^{3+}/Mn^{2+}$ in electron-doped manganites. But the similar rule can also been applied for electron-doped manganites. Fig.6a and Fig.6b displays $T_P$, $T_C$, and $<r_A>$, $\sigma^2$ as a function of the Te doping content, respectively. As for the studied samples $La_{1-x}Te_xMnO_3 (0.1 \leq x \leq 0.6)$, variation of $<r_A>$, $\sigma^2$ and the $Mn^{3+}/Mn^{2+}$ ratio happen simultaneously. As the amount of Te content increases, the carrier density increases and the FM coupling strengthen, accordingly the resistivity would decrease and $T_C$ (or $T_P$) would shift to a higher temperature value. At the same time, $<r_A>$ decreases with the increase of the dopant Te due to the ion size of $Te^{4+}$ is much smaller than that of $La^{3+}$, which would lead to the reduction in $T_C$ (or $T_P$) via modifying the Mn-O-Mn bond distortion. In addition, A-cation radius distribution $\sigma^2$ increases with the increase of Te content accordingly result in the decrease of $T_C$ (or $T_P$). As a result, the competition of the two mechanisms suggested above would decide the change of the Curie temperature $T_C$ and the metal-insulator transition temperature $T_P$. That is why we can see an extremum exhibited during the change of x value. The inset of Fig.5a shows the variation of the Mn-O-Mn bond angle as a function of the Te content. As we can see, the change of Mn-O-Mn bong angle derived by refinement of the structure fit is highly similar to the variation of $T_C$ and $T_P$. It is well known that one of the possible origins of the lattice distortion of perovskite-based structures is the deformation of the $MnO_6$ octahedra originating from Jahn-Teller (JT) effect that is inherent to the high-spin (S=2) $Mn^{3+}$ ions with double degeneracy of the $e_g$ orbital. Obviously, this kind of distortion is directly related to the concentration of $Mn^{3+}$ ions, accordingly correlated with the content of dopant Te. Another possible origin of the lattice distortion is the average ionic radius of the A-site element. As <$r_A$> decreases, the lattice structure have some change and the bending of the Mn-O-Mn bond increases and the bond angle deviates from 180°. The reason mentioned above suggested that why the change of $T_C$ is not monotonous. For the samples of $La_{1-x}Te_xMnO_3 (0.1 \leq x \leq 0.6)$, the variation of the Mn-O-Mn bond angle gives a strong proof to explain the change of the change of the Curie temperature $T_C$ and the metal-insulator transition temperature $T_P$.

It is worth noting that for hole-doped manganites, there exits a critical $<r_A>_L \approx 1.19$ Å, below which the FMM state disappears at the benefit of the AFMI state [27,28]. But for electron-doped manganites $La_{1-x}Te_xMnO_3 (0.1 \leq x \leq 0.6)$, the average A-site size limit to observe the FMM-PMI transition can be pushed down below $<r_A>_L \approx 1.19$ Å. In fact, the average A-site cation radius <$r_A$> reaches 1.07 Å for the sample with x=0.60.

The magnetoresistance (MR) as a function of temperature is plotted in Fig.7. Here the MR is defined as $\Delta\rho/\rho_H = (\rho_0 - \rho_H)/\rho_H$, where $\rho_0$ is the resistivity at zero field and $\rho_H$ is the resistivity at an applied magnetic field of 0.5T. For the sample with x=0.10 and 0.15, there are corresponding peaks at the metal-insulator transition temperature on the MR curves. Moreover, all the samples have evident MR at low temperatures. This behaviour is identical to what was observed in other polycrystalline samples [32, 33].

## 4. Conclusion

In summary, we have studied the effect of Te doping on the structural, electrical transport and magnetic properties of electron-doped manganites system $La_{1-x}Te_xMnO_3 (0.1 \leq x \leq 0.6)$. From these results, the magnetic phase diagram can be proposed for the $La_{1-x}Te_xMnO_3 (0.1 \leq x \leq 0.6)$ manganites, plotting $T_C$ as a function of the $Mn^{2+}$ content (electron carrier density). All samples exhibit paramagnetic-ferromagnetic phase transition accompanied with metal-insulator transition. Three factors including the content of dopant Te, the average A-site cation radius $<r_A>$ and the A-site size mismatch $\sigma^2$ associated to the Mn-O-Mn bond angle have an influence on $T_C$ (or $T_P$). Moreover, the parameter $T_0$ derived by the variable-range hopping fitting data can also reflect the variation of $T_P$. For all the samples, there has an evident MR at low temperatures.

**ACKNOWLEDGMENTS**

This work was supported by the National Key Research under contract No.001CB610604, and the National Nature Science Foundation of China under contract No. 10074066, 10174085, Anhui Province NSF Grant No.03046201 and the Fundamental Bureau Chinese Academy of Sciences.

# References

[1] S. Jin, T. H. Tiefel, M. McCormack, R. A. Fastnacht, R. Ramesh, and L. H. Chen, Science **264**, (1994) 413.

[2] C. Zener, Phys. Rev. **82**, (1951) 403.

[3] A. J. Millis, P.B. littlewood, and B. I. Shraiman, Phys. Rev. Lett. **74**, (1995) 5144.

[4] S. Mori, C. H. Chen, and S-W. Cheong, Phys. Rev. Lett. **81**, (1998) 3972; M. Uehara et al., Nature (London) **399**, (1999) 560.

[5] A. Moreo, S. Yunoki, and E. Dagotto, Science **283**, (1999) 2034.

[6] M. B. Salamon, P. Lin, and S. H. Chun, Phys. Rev. Lett. **88**, (2002) 197203.

[7] P. Mandal and S. Das, Phys. Rev. B **56**, (1997) 15073.

[8] J. R. Gevhardt, S. Roy, and N. Ali, J. Appl. Phys. **85**, (1999) 5390.

[9] P. Raychaudhuri, S. Mukherjee, A. K. Nigam, J. John, U. D. Vaisnav, and R. Pinto, J. Appl. Phys. **86**, (1999) 5718.

[10] J-S Kang, Y J Kim, B W Lee, C G Olson, and B I MIN, J. Phys.: Condens. Matter **13**, (2001) 3779.

[11] Sujoy Roy and Maushad Ali, J. Appl. Phys. **89**, (2001) 7425.

[12] P. Raychaudhuri, C. Mitra, A. Paramekanti, R. Pinto, A. K. Nigam, and S. K. Dhar, J. Phys.: Condens. Matter **10**, (1998) L191.

[13] G. T. Tan, S. Y. Dai, P. Duan, Y. L. Zhou, H. B. Lu, and Z. H. Chen, J. Appl. Phys. **93**, (2003) 5480.

[14] G. T. Tan, P. Duan, S. Y. Dai, Y. L. Zhou, H. B. Lu, and Z. H. Chen, J. Appl. Phys. **93**, (2003) 9920.

[15] G. T. Tan, S. Dai, P. Duan, Y. L. Zhou, H. B. Lu, and Z. H. Chen, Phys. Rev. B **68**, (2003) 014426.

[16] D. B. Wiles and R. A. Young, J. Appl. Crystallogr. **14**, (1981) 149.

[17] Kirby Dwight and Norman Menyuk, Phys. Rev. **119**, (1960) 1470.

[18] A. S. Borovik-Romanov and M. P. Orlova, J. Exptl. Theoret. Phys. (U. S. S. R.) **32**, (1957) 1255.

[19] M. Itoh, T. Shimura, J. D. Yu, T. Hayashi, and Y. Inaguma, Phys. Rev. B **52**,

(1995) 12522.

[20] X. L. Wang, S. J. Kennedy, H. K. Liu, and S. X. Dou, J. Appl. Phys. **83**, (1998) 7177.

[21] N. Zhang, W. P. Ding, W. Zhong, D. Y. Xing, and Y. W. Du, Phys. Rev. B **56**, (1997) 8138.

[22] Ilryong Kim, Joonghoe Dho, and Soonchil Lee, Phys. Rev. B **62**, (2000) 5674.

[23] A. Sundaresan, A. Maignan, and B. Raveau, Phys. Rev. B **56**, (1997) 5092.

[24] J.Fontcuberta, B. Martinez, A. Seffar, S. Pinol, J. L. Garcia-Munoz, and X. Obradors, Phys. Rev. Lett. **76**, (1996) 1122.

[25] Lide M. Rodriguez-Martinez, and J. Paul Attfield, Phys. Rev. B **54**, (1996) R15622.

[26] R. D. Shannon, Acta Crystallogr. Sec. A **32**, (1976) 751.

[27] H. Y. Hwang. S-W. Cheong, P. G. Radaelli, M. Marezio, and B. Batlogg, Phys. Rev. Lett. **75**, (1995) 914.

[28] A. Barnave, A. Maignan, M. Hervieu, F. Damay, C. Martin, and B. Raveau, Appl. Phys. Lett. **71**, (1997) 3907,.

[29] L. Sheng, D. N. Sheng, and C. S. Ting, Phys. Rev. B **59**, (1999) 13550.

[30] Lide M. Rodriguez-Martinez, and J. Paul Attfield, Phys. Rev. B **58**, (1998) 2426.

[31] P. V. Vanitha, P. N. Santhosh, R. S. Singh, C. N. Rao, and J. P. Attfield, Phys. Rev. B **59**, (1999) 13539.

[32] Hongsuk Yi and Jaejun Yu, Phys. Rev. B **58**, (1998) 11123.

[33] Lei Zheng, Xiaojun Xu, Li Pi, and Yuheng Zhang, Phys. Rev. B **62**, (2000) 1193.

# Tables:

TABLE I. Room-temperature structural parameters and the fitting parameter $T_0$ of $La_{1-x}Te_xMnO_3 (0.1 \leq x \leq 0.6)$ samples.

| x | a (Å) | c (Å) | V (Å$^3$) | θ$_{Mn-O-Mn}$ (°) | d$_{Mn-O}$ (Å) | $T_0$ |
|---|---|---|---|---|---|---|
| 0.10 | 5.524 | 13.357 | 353.1012 | 163.81 | 1.9645 | $2.95 \times 10^7$ |
| 0.15 | 5.525 | 13.355 | 353.0998 | 163.83 | 1.9644 | $2.36 \times 10^7$ |
| 0.20 | 5.529 | 13.358 | 353.5864 | 162.24 | 1.9694 | $1.54 \times 10^8$ |
| 0.25 | 5.524 | 13.490 | 356.5293 | 159.10 | 1.9840 | $2.13 \times 10^8$ |
| 0.30 | 5.541 | 13.470 | 357.8054 | 160.72 | 1.9806 | $1.76 \times 10^8$ |
| 0.35 | 5.528 | 13.358 | 353.4509 | 162.88 | 1.9675 | $1.65 \times 10^8$ |
| 0.40 | 5.530 | 13.512 | 357.8558 | 161.35 | 1.9798 | $1.54 \times 10^8$ |
| 0.45 | 5.548 | 13.409 | 356.9382 | 159.12 | 1.9835 | $1.72 \times 10^8$ |
| 0.50 | 5.540 | 13.427 | 356.5127 | 161.34 | 1.9764 | $1.47 \times 10^8$ |
| 0.60 | 5.537 | 13.428 | 356.2108 | 161.58 | 1.9721 | $1.45 \times 10^8$ |

# Figure captions:

Fig.1. The magnetic diagram of the compound $La_{1-x}Te_xMnO_3$ $(0.1 \leq x \leq 0.6)$, based on the present work. The Curie temperature was derived by the M-T curve.

Fig.2. XRD patterns of the compound $La_{1-x}Te_xMnO_3$ $(x = 0.10)$. Crosses indicate the experimental data and the calculated data is the continuous line overlying them. The lowest curve shows the difference between experimental and calculated patterns. The vertical bars indicate the expected reflection positions.

Fig.3. (a) The temperature dependence of magnetization in $La_{1-x}Te_xMnO_3$ $(0.1 \leq x \leq 0.6)$ measured in a magnetic field of 0.1T. Field cooling curves are shown. (b) The frequency dependence of the ac magnetic susceptibility ($\chi$) for the samples with x=0.20 and 0.40. The inset shows the temperature dependence of the FC and ZFC response of the sample with x=0.20 measured in external field of 0.1T.

Fig.4. (a)The temperature dependence of the resistivity of $La_{1-x}Te_xMnO_3$ $(0.1 \leq x \leq 0.5)$ samples in zero and 0.5T fields. (b) The temperature dependence of the resistivity with the sample with x=0.60 in zero and 0.5T fields. The inset by the arrow direction is the magnified plot at low temperature. The inset below shows the magnetization as a function of the applied magnetic field at 5K.

Fig.5. The varible-range hopping fitting of resistivity curves of $La_{1-x}Te_xMnO_3$ $(0.1 \leq x \leq 0.6)$. The dashed lines represented the experimental data.

Fig.6. (a) The variation of $T_C$ and $T_P$ with the Te content. The inset is the Mn-O-Mn

bond angle dependence of the Te content. (b) The variation of the average A-site cation radius <$r_A$> and the A-site cation mismatch $\sigma^2$ with the dopant Te.

Fig.7. The temperature dependence of magnetoresistance (MR) of $La_{1-x}Te_xMnO_3 (0.1 \leq x \leq 0.6)$. The direction of arrow denotes the sequence of the Te doping level.

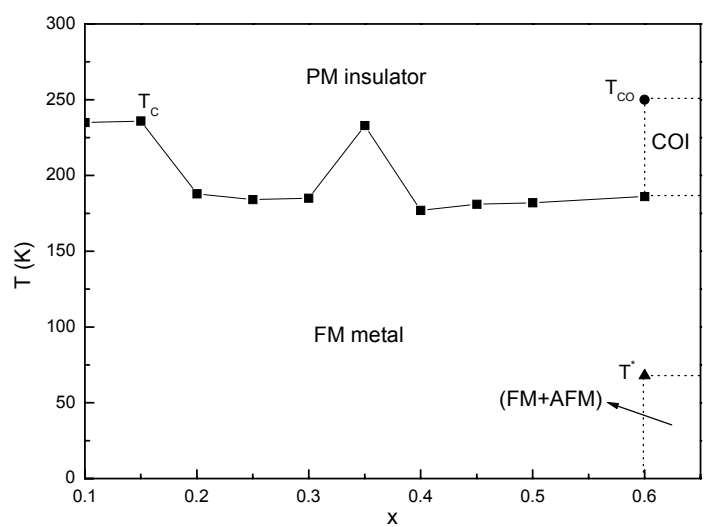

Fig.1. J. Yang et al.

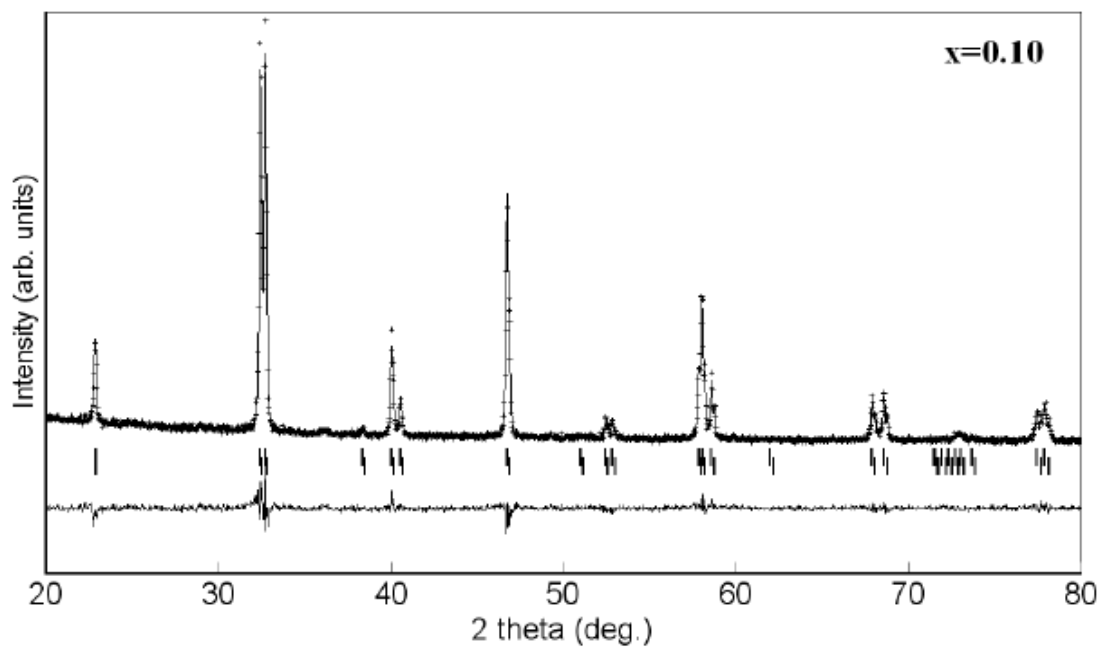

Fig.2. J. Yang et al.

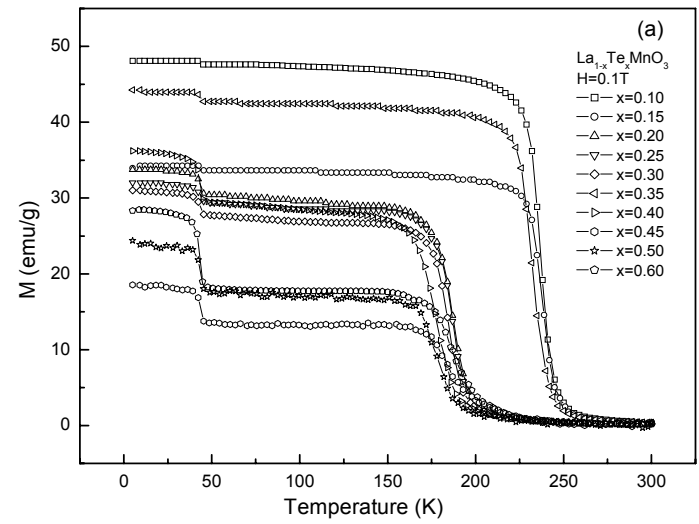

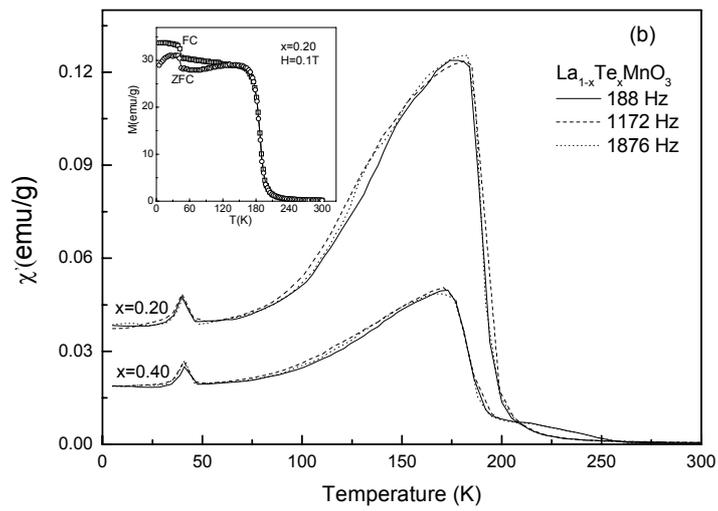

Fig.3. J. Yang et al.

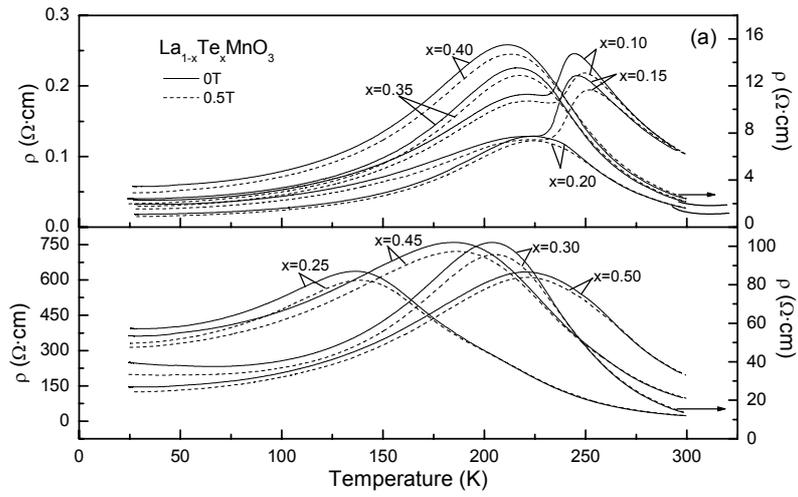

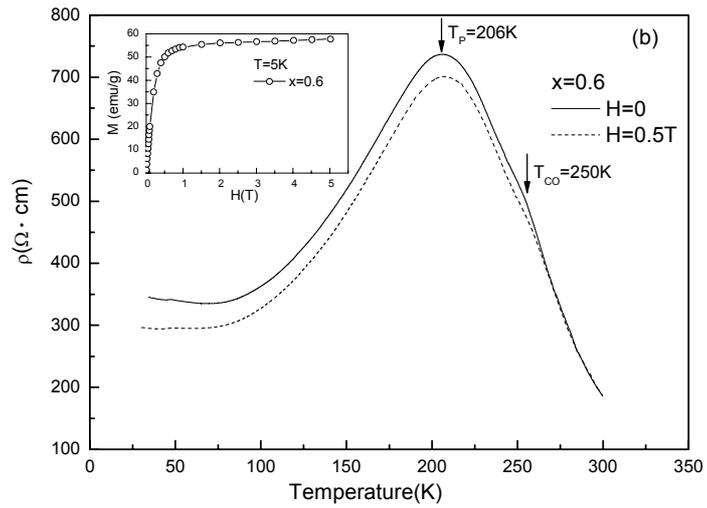

Fig.4. J. Yang et al.

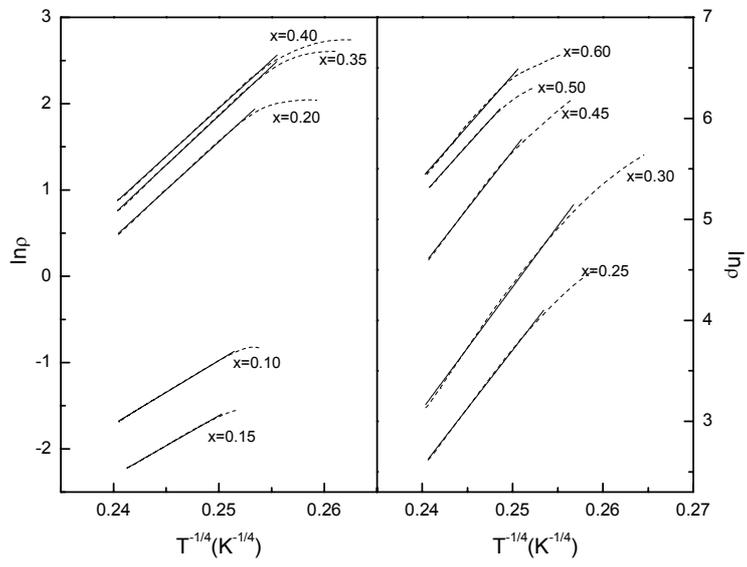

Fig.5. J. Yang et al.

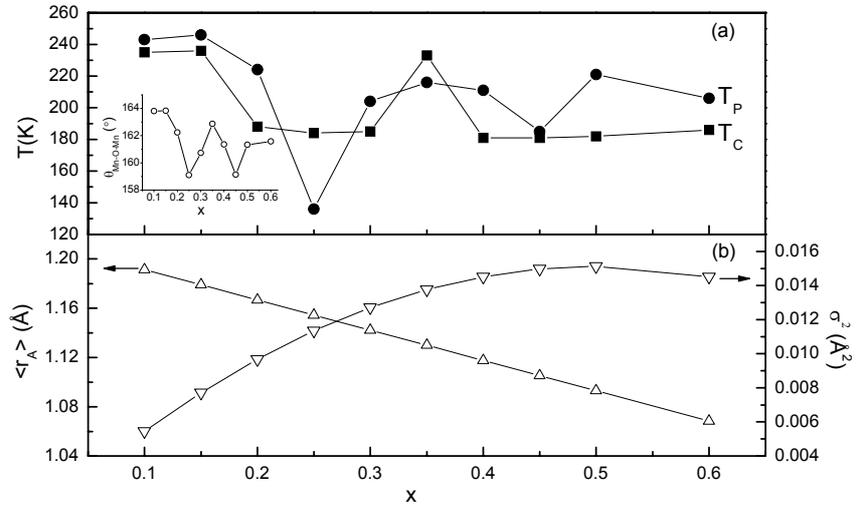

Fig.6. J. Yang et al.

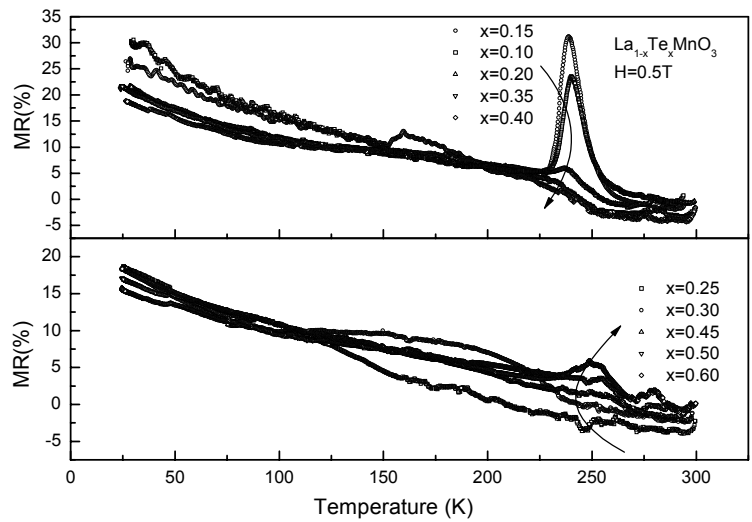

Fig.7. J. Yang et al.